\documentclass{pasj00}
\usepackage{natbib}
\usepackage{mnsup}
\begin{document}

\SetRunningHead{R. Ishioka}{HT Cam}

\title{Dwarf Nova-like Outburst of Short Period Intermediate Polar\\
 HT Camelopardalis}
\author{Ryoko \textsc{Ishioka}, Taichi \textsc{Kato}, Makoto
\textsc{Uemura}}
\affil{Department of Astronomy, Faculty of Science, Kyoto University,
       Sakyou-ku, Kyoto 606-8502, Japan}
\email{ishioka@kusastro.kyoto-u.ac.jp, tkato@kusastro.kyoto-u.ac.jp,
uemura@kusastro.kyoto-u.ac.jp,}
\email{bemart1@telusplanet.net}
\author{Gary W. \textsc{Billings}}
\affil{2320 Cherokee Dr NW, Calgary, Alberta, Canada, T2L 0X7}
\email{obs681@telusplanet.net}
\author{Koichi \textsc{Morikawa}}
\affil{VSOLJ, 468-3 Satoyamada, Yakage-cho, Oda-gun, Okayama 714-1213, Japan}
\author{Ken'ichi \textsc{Torii}}
\affil{Cosmic Radiation Laboratory, Institute of Physical and Chemical Research (RIKEN),\\
 2-1, Wako, Saitama, 351-0198, Japan} 
\author{Kenji \textsc{Tanabe}}
\affil{Department of Biosphere-Geosphere Systems, Faculty of Informatics, Okayama University of Science,\\
Ridaicho 1-1, Okayama 700-0005, Japan}
\author{Arto \textsc{Oksanen}, Harri \textsc{Hyv\"{o}nen}}
\affil{Nyr\"{o}l\"{a} Observatory, Jyv\"{a}skyl\"{a}n Sirius ry,
 Kyllikinkatu 1, FIN-40100  Jyv\"{a}skyl\"{a}, Finland}
\author{Hitoshi \textsc{Itoh}}
\affil{VSOLJ, Nishiteragata-cho 1001-105, Hachioji, Tokyo 192-0153, Japan}
\Received{}
\Accepted{}

\maketitle

\begin{abstract}
   We report the first time-series observations of the short outburst of
 the proposed intermediate polar HT Cam (=RX J0757.0+6306). On 2001
 December 29, we detected the object was undergoing a bright outburst
 at the magnitude of $m_{vis}=12.2$. Following this detection, we started 
 international joint observations through VSNET. The light curve showed a
 gradual decline for the first 0.5~d. Following this short plateau phase, the
 rate of decline dramatically increased to more
 than 4~mag d$^{-1}$. Within 1.5~d from the outburst detection, the
 object almost declined to the quiescent level. During the rapidly
 declining phase, long-term modulations with 
 a period of 86~min and strong pulses with a period of 8.6~min were
 observed. We concluded that 86~min and 8.6~min are the orbital period
 and the spin period of HT Cam, respectively. By the detection of the
 spin period, we 
 confirmed the IP classification of HT Cam. However, its outburst
 behavior rather resembles that of dwarf novae. 
  The discrepancy between the declining rates of the total flux and the
 pulse flux strongly suggests that the disk instabilities were taking
 place during the outburst.
 \end{abstract}

\section{Introduction}

Cataclysmic variables consist of a white dwarf primary and 
a late-type main-sequence companion star which is mass-losing by the
Roche-lobe overflow. The white dwarf accretes the mass from the secondary
through an accretion disk in non-magnetic systems.
In magnetic systems, however, either the formation of the disk is fully
prevented, or the inner disk is truncated, depending on the magnetic fields
of the white dwarf and the binary parameters. Within the magnetosphere
radius the mass falls  
along the magnetic field toward poles of the white dwarf, where radiations
from X-ray to optical emission is originated. 
Magnetic systems are divided into two classes of polars and intermediate
polars (IPs). The polars are the systems with strong fields and a
synchronously rotating white dwarf, and are soft X-ray sources. IPs
are ones with weaker fields and an asynchronously rotating white dwarf,
and are hard X-ray sources. Reviews for the magnetic systems are
given by \citet{cro90polar}, \citet{pat94ipreview}, \citet{war95book},
and \citet{hel96IPreview}. 

There are six IPs in which outbursts are observed: GK Per, EX Hya, XY 
Ari, DO Dra\footnote{There has been some confusion regarding the
nomenclature of this object. Some authors call the same variable as YY
Dra. We consistently use the name DO Dra, following the original
nomenclature by the General Catalogue of Variable Stars (GCVS)
\citep{NameList67,kho88dodra}.}, TV Col, and V1223 Sgr.  
GK Per shows long outbursts that last for 2~months
\citep{sab83gkper,kim92gkper}. Remaining 5 IPs show rather short 
outbursts. Outbursts of XY Ari \citep{hel97xyarioutburst} and DO Dra \citep{wen83dodra,szk02dodra} are with amplitudes of $3-5$~mag and durations
of $\sim 5$~d. TV Col \citep{szk84tvcolflare,hel93tvcolperiods} and 
V1223 Sgr \citep{vaname89v1223sgr} show very short
low-amplitude outbursts that last only $\sim 0.5$~d. Outbursts of EX Hya 
have intermediate  characteristics between the previous two types, that
is, with amplitudes of $\sim 3.5$~mag and durations of $2-3$~d
\citep{hel89exhya}.   

Outbursts of GK Per are explained as inside-out outbursts by thermal
instability in the disk, though the atypically large inner radius and
mass transfer rate are required to fit the observed recurrence time and
duration of outbursts \citep{kim92gkper}. 
Outbursts of XY Ari and DO Dra rather resemble 
outbursts of dwarf novae and also explained as inside-out outbursts by
disk instability \citep{hel97xyarioutburst}. 
The recurrence time one order longer than the typical value is thought to be 
due to inner truncation of the disk \citep{ang89DNoutburstmagnetic}. On the
other hand, the short-lived and low-amplitude outbursts of TV Col and
V1223 Sgr are thought to be mass transfer outbursts
\citep{hel93tvcol}. \citet{hel00exhyaoutburst} suggested that outbursts 
of EX Hya are also mass transfer events. 
 
HT Cam is a cataclysmic variable identified as the optical counterpart of 
the hard X-ray source RX~J0757.0$+$6306, discovered during the ROSAT All-Sky 
Survey. Tovmassian et al. (\citeyear{tov97htcam},\citeyear{tov98htcam})
suggested that this object is an intermediate polar with a shortest
orbital period of 80.92 min and a spin period of 8.52 min. However, the
existence of dwarf nova-like outbursts and the short orbital period
allowed an alternative interpretation that it may be an SU UMa-type
dwarf nova or WZ Sge-type stars \citep{tov98htcam}. After 
the discovery, five outbursts were recorded (vsnet-alert 1379, 3025,
4845, 5255, 6066, and \cite{wat98htcam}). All of them were very short and
only one or a few point observations were obtained for each one. 

Here we report the first time-series observations during
its short outburst.

\section{Observations}

The 6th outburst was detected by H. Itoh on 29 December 2001 
(vsnet-alert 6944). The object was at the visual magnitude of 12.2, which is 
6 mag brighter than its quiescence level. We performed time-series CCD 
observations totally on 6 nights at six sites: Department of Astronomy,
Kyoto University, Japan (``Kyoto'' in Table 1), Okayama, Japan 
(``Okayama1''and ``Okayama2'' in Table 1), Cosmic Radiation Laboratory,
Institute of Physical and Chemical Research (RIKEN), Saitama, Japan
(``RIKEN'' in Table 1), Nyr\"{o}l\"{a} Observatory, Jyv\"{a}skyl\"{a},
Finland (``Nyrola'' in Table 1), and Calgary, Alberta, Canada
(``Canada'' in Table 1). A journal of observations and a description  
of the equipment of CCD photometry are given in tables
\ref{tab:obs_log} and \ref{tab:equipment}.

\begin{table*}
\caption{Journal of CCD photometry.} \label{tab:obs_log}
\begin{center}
\begin{tabular}{ccccccc}
\hline
\multicolumn{3}{c}{Date}& Start - End(HJD$-$2450000) & T  & N & Observatory \\
\hline
2001  & December & 30 &2273.62148 - 2274.11198 & 120  &283& Canada \\
      & December & 30 &2273.87041 - 2274.37440 & 30  &544& Kyoto \\
      & December & 30 &2273.96269 - 2274.19700 & 30  &391& Okayama1 \\
      & December & 30 &2274.00834 - 2274.37813 & 30  &777& RIKEN \\
      & December & 30 &2274.01264 - 2274.26793 &12-20&255& Okayama2 \\
      & December & 30 &2274.42271 - 2274.44218 & 60  & 23& Nyrola \\
      & December & 31 &2274.87649 - 2275.23216 & 30  &548& Kyoto  \\
      & December & 31 &2274.99203 - 2275.45142 & 30  &863& RIKEN \\
2002  & January  & 1  &2275.98462 - 2276.18404 & 30  &414& Okayama1 \\
      & January  & 1  &2276.00891 - 2276.31457 & 30  &662& RIKEN \\
      & January  & 2  &2277.00627 - 2277.50296 & 30  &674& RIKEN \\
      & January  & 3  &2278.02761 - 2278.30901 & 30  &603& RIKEN \\
      & January  & 4  &2279.01785 - 2279.25500 & 30  &508& RIKEN \\
\hline
\end{tabular}
\end{center}
\end{table*}

\begin{table}
\caption{Equipment of CCD photometry.} \label{tab:equipment}
\begin{center}
\begin{tabular}{cccc}
\hline
Observer       &  Telescope      &  CCD  & Comparison \\ 
\hline
Kyoto      & 25 cm SC        & ST-7E & GSC 4117:827$^{a}$ \\
RIKEN       & 20 cm reflector & AP-7p    & GSC 4117:827$^{a}$ \\
Okayama1      & 30 cm C         & ST-9E  & GSC 4117:827$^{a}$ \\
Okayama2  &  25 cm SC   &  ST-7   & GSC 4117:631$^{b}$  \\
Canada    &  36 cm SC   &  V-filtered ST-7E &   GSC 4117:827$^{a}$ \\
Nyrola    &   40 cm SC  & ST-8E  & GSC 4117:1267$^{c}$ \\
\hline
\multicolumn{4}{l}{a. 075644.8+630742 V$=$10.91 B$-$V$=$0.47 (Henden)}\\
\multicolumn{4}{l}{b. 075650.9+630521 V$=$11.45 B$-$V$=$0.83 (Henden)}\\
\multicolumn{4}{l}{c. 075711.9+630720 R$=$13.20 B$=$14.46 (GSC-II)}\\
\end{tabular}
\end{center}
\end{table}

After correcting for the standard de-biasing and flat fielding, we
processed object frames with the PSF and aperture photometry packages.
We performed differential photometry relative to the comparison stars 
listed in table \ref{tab:obs_log}, whose constancy was confirmed between
each other. Finally, a heliocentric corrections was applied before the
following analysis.

\section{Results}

\begin{figure}[]
  \begin{center}
  \FigureFile(80mm,80mm){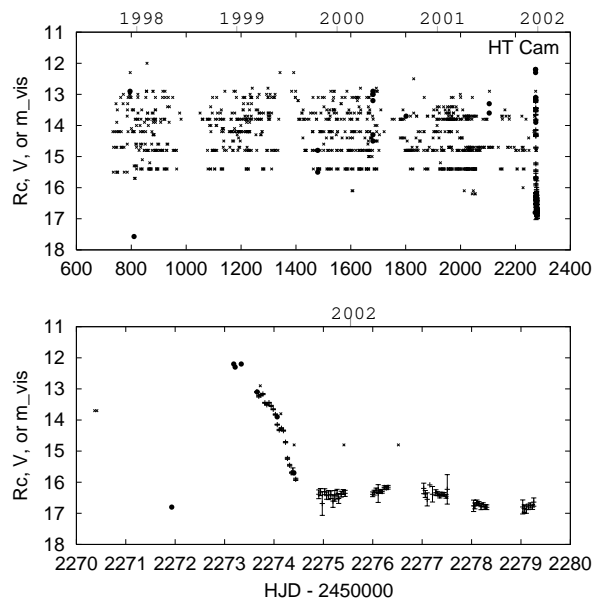}
  \end{center}
  \caption{Long-term light curves of HT Cam. Abscissa and ordinate
 denote HJD (also calendar date) and $R_{c}$, $V$, or visual magnitude,
 respectively. Crosses with error bar show average magnitude in 1-hr bin of the
 time-series CCD observations. Filled circles and small crosses show
 positive detections by visual or snapshot CCD observations and
 upper-limits of visual observations reported to Variable Star NETwork
 (VSNET). Upper panel: light curve from 1997 October to 2002 January. Lower
 panel: light curve of the outburst on 2001 December.}\label{fig:general} 
\end{figure}

Figure \ref{fig:general} shows long-term light curves of HT Cam.  The
upper panel is the 
light curve including all visual and CCD observations reported to
Variable Star NETwork
(VSNET\footnote{http://www.kusastro.kyoto-u.ac.jp/vsnet/}). The lower panel is
the enlarged light curve of the outburst in 2001 December. 
Filled circles show positive detections by visual or snapshot CCD
observations.  Crosses with error bar show average magnitude in 1-hr bin of the
time-series CCD observations. Small crosses show upper-limits by visual
observations. 

\begin{figure}[]
  \begin{center}
  \FigureFile(80mm,80mm){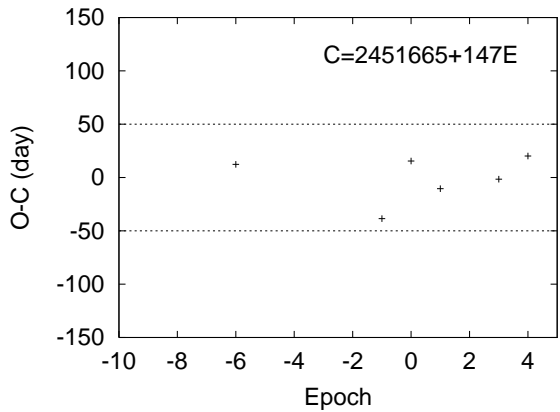}
  \end{center}
  \caption{O-C diagram for the moment of the six outbursts detected at
 HJD 2450795, 2451479, 2451680, 2451801, 2452104, and 2452273. E$=$0
 corresponds to the outburst at HJD~2451680.}\label{fig:o-c}  
\end{figure}

Including this outburst, 6 outbursts were observed in 4 years. However
the observations are not dense or deep enough to catch all the short (3-d)
outbursts ($V_{max} \sim 12$~mag), so we may have missed several ones.
We searched the recurrence time of the outbursts in HT Cam from the 6
outbursts, assuming regularity and that we may have missed several ones, and
found a fundamental period of $\sim 150$~d. Figure \ref{fig:o-c} shows
the $O-C$ diagram  
for this period. The deviation of  $|O-C|$s is within $\sim 50$~day, one 
third of the period we obtained. 

The past five outbursts were only by one or few points visual
observations, since the outbursts were very short as those of other
intermediate polars except GK Per. The 6th outburst in 2001 December
detected at 0.7 mag brighter magnitude than previous outbursts. This
indicates the detection at near the outburst maximum. 

As shown in the lower panel of figure \ref{fig:general}, we started time-series
observation 0.5~d after the outburst detection and succeeded in
observing during the declining phase of the short outburst. The rising
phase was missed, but the last quiescent observation reported to VSNET
was performed within 1.5~d before the detection. Thus the rising rate is 
faster than 3~mag~d$^{-1}$.
On the first night, the brightness declined very rapidly, 2.5 mag within
0.75~d. The declining rate changed at HJD2452274.0 from 1.5~mag~d$^{-1}$
to 5.5~mag~d$^{-1}$. The object already declined to nearly its quiescent
level, however 0.3~mag brighter than quiescence, 1.5~d after the
detection of the outburst. It stayed almost the same state for three
days. The object returned to its quiescent level of $\sim 16.8$~mag 5~d
after the outburst detection. 

\begin{figure*}[]
  \begin{center}
  \FigureFile(160mm,80mm){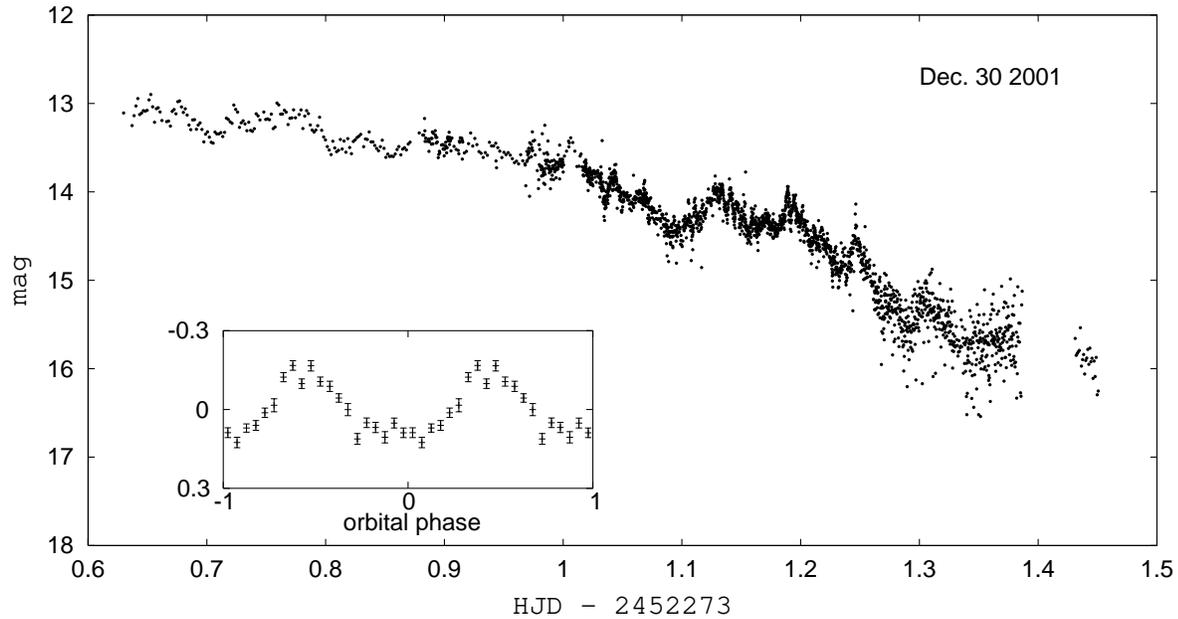}
  \end{center}
  \caption{Light curve of time-series observations in first night,
 during the declining phase from $\sim 13.1$ to $\sim 15.9$ with a rate
 of $\sim 4.8$ mag day$^{-1}$. Abscissa and ordinate denote HJD and
 magnitude, respectively. The light curve shows
 quasi-periodic  oscillations with a typical amplitude of 0.5 mag and a
 time scale close to the orbital period. The small panel shows an
 averaged plofile of the orbital modulations. The abscissa and ordinate
 denote orbital phase and differential magnitude.}\label{fig:1st_night}   
\end{figure*}

\begin{figure*}[]
  \begin{center}
  \FigureFile(160mm,80mm){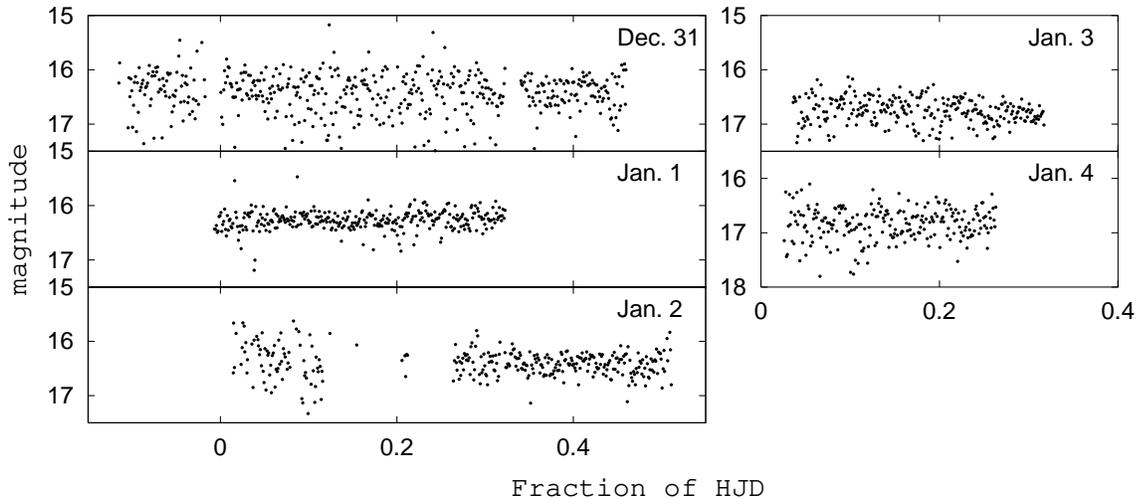}
  \end{center}
  \caption{Nightly light curves of time-series observations from 2001
 Dec. 31 to 2001 Jan. 4, after the object declined to quiescence with
 magnitude range of $R_{c} = 16 - 17$. The linear trend was subtracted
 from each light curve and each point is an average of 90-s bin. Abscissa
 and ordinate denote fraction of HJD and differential magnitude, 
 respectively. Dec. 31 is HJD 2452275. Light curves show no modulation 
 with orbital period.}
 \label{fig:nightly_lc}   
\end{figure*}

Figure \ref{fig:1st_night} shows a light curve on Dec. 30,  
that is during the declining phase. This light curve shows
quasi-periodic modulations with a time scale of $\sim 0.06$~d, close to
the orbital period, superimposed on long-term variations. The small
panel shows an averaged profile of the orbital modulations. We cannot detect 
the modulations with the orbital period in the light curves after the
object declined to quiescence (figure \ref{fig:nightly_lc}).

\begin{figure}[]
  \begin{center}
  \FigureFile(80mm,80mm){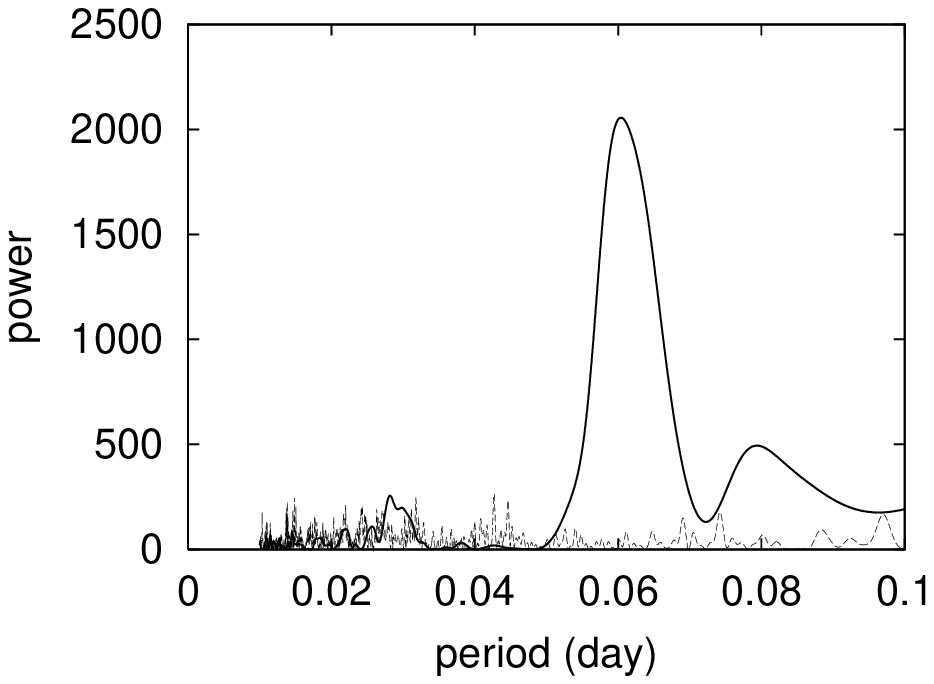}
  \end{center}
  \caption{Power spectra. The abscissa is the period in day and the
 ordinate is the power in arbitrary unit. The thick line and the thin
 broken-line show the results for the data on Dec. 30 and
 Dec. 31--Jan. 4, respectively. The diagram show strong signal with
 wide widths with a center period of $\sim 0.060$ min on Dec. 30.}\label{fig:long_period}     
\end{figure}

We applied a period search by Fourier transform and phase dispersion 
minimization (PDM) methods with period ranges of 0.01--0.1~d to the data
of Dec. 30 in magnitude scale and combined data of Dec. 31 -- Jan. 4 in
flux scale. Figure \ref{fig:long_period} 
exhibits the resultant power spectra. The spectrum on the data of
Dec. 30 shows strong peak with wide width around the period of $\sim
0.06$~d. We determined the center period to be $0.0606 \pm 0.0004$~d 
($87.2 \pm 0.6$ min) by fitting Gaussian function. 
In contrast, the spectrum for the data of Dec. 31 -- Jan. 4
show no strong signal. PDM method gives consistent results.

\begin{figure}[]
  \begin{center}
  \FigureFile(80mm,80mm){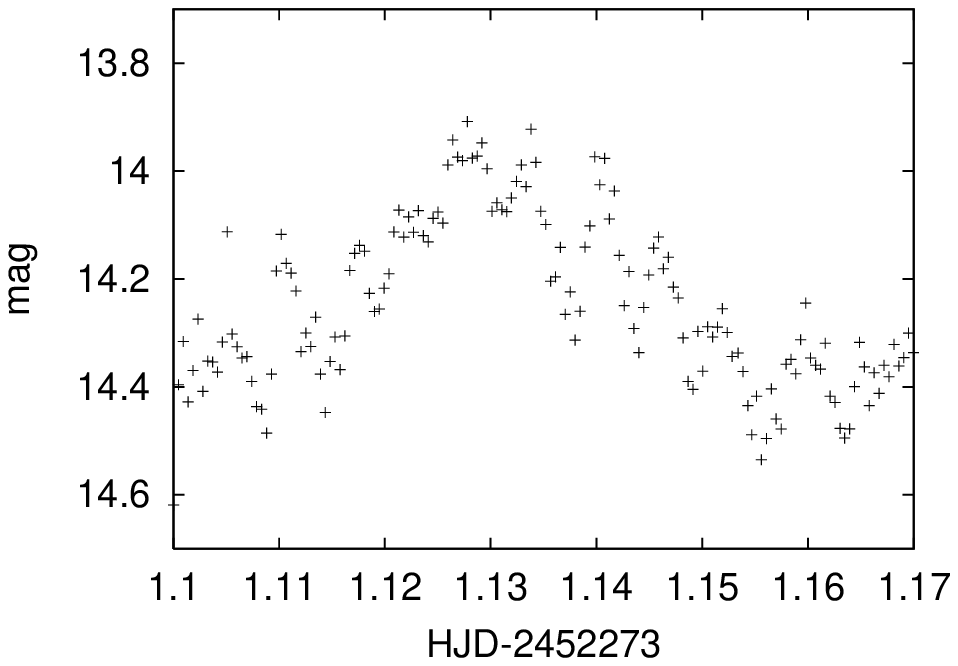}
  \end{center}
  \caption{Enlarged light curves which shown in figure
 \ref{fig:1st_night}. Abscissa and ordinate denote HJD and differential
 magnitude, respectively. Quasi-periodic oscillations with short periods 
 are superimposed on the orbital modulations. The typical value of the error is
 $\sim 0.05$ mag.}\label{fig:pulse}  
\end{figure}

\begin{figure*}[]
  \begin{center}
  \FigureFile(160mm,80mm){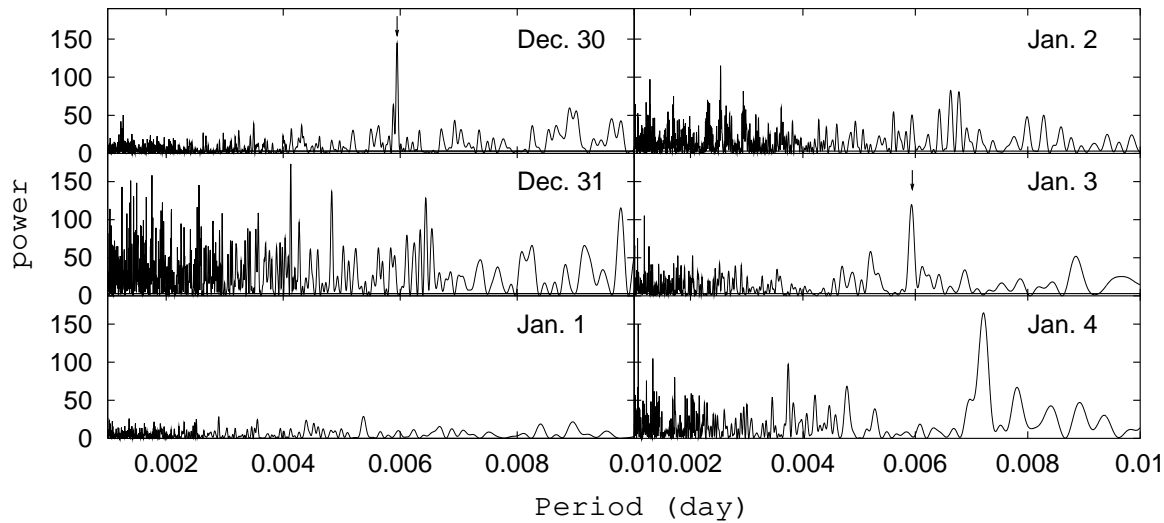}
  \end{center}
  \caption{Power spectra on the nightly light curves. The abscissa and
 the ordinate of each panel are the period in day and the power in the
 arbitrary unit. The top panel and the fifth panel show strong peaks at
 the period of 0.005946 d (marked by small arrows).}\label{fig:nightly_fou}
\end{figure*}

In addition to the modulations with the orbital period, our light curves 
show coherent pluses with a short period (figure \ref{fig:pulse}).
Figure \ref{fig:nightly_fou} shows power spectra with period range of
0.001--0.01~d for 
nightly light curves shown in figure \ref{fig:nightly_lc}, though the
light curve on 2001 Dec. 30 
is that of after subtracting the modulations with a typical time
scale of 0.06~d. The abscissa and the ordinate of each panel are
the period in day and the power in arbitrary unit, respectively. The top 
panel shows a strong peak at the period of $0.005946 \pm 0.000003$ d
($8.562 \pm 0.004$ min). Next three panels show no strong peak at this 
period, however week peaks exist at the close periods. The fifth panel
from the top shows a strong peak at the same period of $0.005946 \pm
0.000001$~d ($8.562 \pm 0.001$ min). The bottom panel shows a wide-width
peak at about 0.0072~d.  

\begin{figure}[]
  \begin{center}
  \FigureFile(80mm,80mm){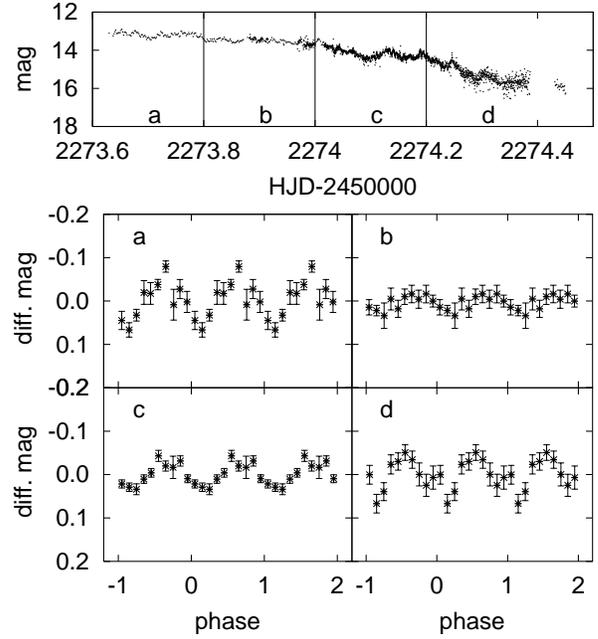}
  \end{center}
  \caption{Averaged spin profiles during the declining
 phase. The declining rate changed at HJD 2452274.0.}\label{fig:spin_profile} 
\end{figure}

Figure \ref{fig:spin_profile} shows averaged spin profiles of the data
on Dec. 30, after 
subtracting the modulations with a typical time of $\sim 0.06$ d and
divided into four parts represented in the top panel. At first the
amplitude of the spin pulse was $\sim 0.2 \rm{mag}$~(a), but it
decreased to $\sim 0.1\rm{mag}$~(b). Then it gradually increased again during
the rapidly declining phase (c,d).

\section{Discussion}

\subsection{8.6 min oscillation} 

Short-term coherent pluses with a period of 8.6~min were observed during the
outburst. This period is thought to be the spin period of white
dwarf. \citet{tov98htcam} found a period of 8.5~min from their
quiescence observations. We could not detect the 8.6~min period after
the object declined except on Jan. 3. The pulse amplitude should have
been below the limit of detection, $\sim 0.2$~mag. 

The pulse amplitude in flux scale continuously reduced as the system
declined, but in magnitude scale it decreased once and increased
again during the rapidly declining phase (figure
\ref{fig:spin_profile}), that is, the 
declining rate of the pulse flux was slower than that of the total flux
of the system. The pulse flux and the total flux roughly
correspond to the accretion rate onto the white dwarf and the
accretion rate through the outer region. The discrepancy of declining
speeds of two accretion rates suggests the existence of a disk, rather
than via ballistic accretion.  

\subsection{86 min oscillation}

\citet{tov98htcam} suggested 81~min as the orbital period of HT Cam
from the radial velocity curves, but they could not reject 76~min and
86~min. Their quiescent photometry showed no modulations with the period
of $\sim$81~min. This indicates the low inclination of HT Cam.
In our observations, after the declining to quiescence, no modulations
with the orbital period were observed, however, large-amplitude
quasi-periodic modulations with the period of 86~min were observed
during the declining phase of the outburst. These modulations are
likely related to the orbital period.  

If the orbital period of HT Cam is 81~min, our period is longer than it.
During superoutbursts in SU UMa-type dwarf novae,
modulations with a period a few percent longer than its
orbital period, called superhumps are observed \citep{war85suuma}. It is known that
superhumps are detected 
regardless of the inclination of the system. However, in conjunction with the
short duration of the outburst, we consider the 86-min unlikely
represents the superhump period.  This period more likely reflects
the true orbital period among the aliases listed in \citet{tov98htcam}. 

In WZ Sge stars, a small sub-group of SU UMa-type dwarf novae with very short
orbital periods of around 80--90~min, modulations with orbital periods,
called early superhumps, appear in the early stage of superoutbursts
\citep{mat98egcnc}. Early superhumps
have double-humped plofile regardless of the inclination of
the system. However, the averaged profile shown in figure
\ref{fig:1st_night} is not double-humped. Thus the possibility that the
modulations were early superhumps is excluded. 

The existence of the orbital modulations only during the outburst may be 
due to the changing aspect of a irradiated surface of the secondary star,
an enhanced hot spot moving into and out of our line of sight, partial
eclipse of the thickend and brightened disk, or a combination of them.

\subsection{Long-term variability}

The visual observations covering four years with fairly good density have
detected 6 rather regular outbursts with an amplitude of $\sim 4$~mag and a
duration of $< 3$~d. Our observations on the 6th
outburst of HT Cam confirmed the short duration and rapid fading,
perhaps slightly slower than rising. The outburst amplitude was $\sim
4.5$~mag, though its duration was very short. 

EX Hya shows outbursts with a similar short duration of $\sim 3$~d and
the amplitude of $\sim 3$~mag.
\citet{hel00exhyaoutburst} mentioned that the outbursts of EX Hya 
are mass transfer events rather than disk instabilities, due to the
rarity and irregularity of the outbursts (e.g. no outburst was
observed in a period of more than 10 yr, however, in another period, two
outbursts occurred with an interval of only 8~d), the declining rate, the
increased equivalent widths of emission lines and the enhanced
mass-transfer stream during the outbursts.

However,
the outburst profile, which shows rather slower declining at first
and then very fast declining later, and the regularity of the outburst
recurrence time of HT Cam are similar to those of DO Dra. 
DO Dra shows outbursts with a duration of $\sim 5$~days and a recurrence time
of 868~d \citep{sim00dodra}. They are explained by disk instability as dwarf
novae \citep{szk02dodra}. 

The differences of HT Cam from DO Dra are the shorter outburst
duration within $3$~d, and the shorter cycle length of $\sim 150$~d. 
The recurrence time of HT Cam is quite typical value among 
the dwarf novae, compared with those of two intermediate polars 
whose outbursts are likely explained by disk instabilities (
$\sim 870$~d of DO Dra as written above, and $\sim 900$~d or more of GK
Per; \cite{sim02gkper}).   
There are several dwarf novae that have recurrence times about or longer
than $\sim 150$~d and show short outbursts about 3--5~d (e.g. HT Cas, EK
TrA, IR Com). 

\citet{ang89DNoutburstmagnetic} calculated the effects of the inner disk
truncation on the outbursts in the intermediate polars:
the duration becomes shorter in both cases of mass transfer
instability and disk instabilities, and for the outbursts by disk
instabilities, the interval becomes longer if the outburst starts at the
inside, however does not change if the outburst starts at the outer disk 
edge. 

Using this result, we can explain the outburst behavior of HT Cam
as follows. The outbursts are caused by disk instabilities
starting at the outside, thus the recurrence time is comparable to those
of dwarf novae, $\sim 150$~d. The short duration and the rapid decline
of the outburst are due to the inner truncation of the disk. 
This is consistent with the rather short spin period indicating
intermediate magnetic fields \citep{kin91IPspinevolution}, and the
discrepancy between the declining rates of pulse flux and total flux
strongly suggesting the existence of the disk.

\section{summary}

We observed the cataclysmic variable HT Cam in outburst which occurred
on 2001 December 29. We detected quasi-periodic oscillations in the
light curve of the declining phase of the outburst whose center periods
are calculated to be 8.56~min and 86.5~min. Strong 8.56~min oscillations
during the outburst are white dwarf spin pulses and the IP classification
was confirmed. We identified the 86.5~min period as the 
orbital period of HT Cam, which was determined as $81 \pm 5$~min by
Tovmassian et al. (1998). The rather short spin period indicates the
magnetic field of HT Cam is not so strong, and the discrepancy between
the declining rates of pulse flux and total flux suggests the existence
of the disk.  Additionally, we found that the intervals of the past six
outbursts are the multiples of 150~d. The regularity of outbursts
suggests that the outbursts on HT Cam is caused by disk instabilities
starting at the outer disk edge.  

\medskip

We are grateful to many VSNET observers who have reported vital observations.
We also thank D. Nogami for his comments. The CCD camera of AO and HH is
on loan from AAVSO. GWB acknowledges the use of software provided by the
Starlink Project which is funded by the UK Particle Physics and
Astronomy Research Council. This work is partly supported by a grant-in aid
(13640239) from the Japanese Ministry of Education, Culture, Sports,
Science and Technology. Part of this work is supported by a Research
Fellowship of the Japan Society for the Promotion of Science for Young
Scientists (MU).

\end{document}